# 4MOST Consortium Survey 10: The Time-Domain Extragalactic Survey (TiDES)

Elizabeth Swann[1]
Mark Sullivan[2]
Jonathan Carrick[3]
Sebastian Hoenig[2]
Isobel Hook[3]
Rubina Kotak[4,5]
Kate Maguire[4]
Richard McMahon[6]
Robert Nichol[1]
Stephen Smartt[4]

[1] Institute of Cosmology and Gravitation, University of Portsmouth, UK
[2] School of Physics and Astronomy, University of Southampton, UK
[3] Physics Department, Lancaster University, UK
[4] School of Mathematics and Physics, Queen's University Belfast, UK
[5] University of Turku, Finland
[6] Institute of Astronomy, University of Cambridge, UK

The Time-Domain Extragalactic Survey (TiDES) is focused on the spectroscopic follow-up of extragalactic optical transients and variable sources selected from forthcoming large sky surveys such as that from the Large Synoptic Survey Telescope (LSST). TiDES contains three sub-surveys: (i) spectroscopic observations of supernova-like transients; (ii) comprehensive follow-up of transient host galaxies to obtain redshift measurements for cosmological applications; and (iii) repeat spectroscopic observations to enable the reverberation mapping of active galactic nuclei. Our simulations predict we will be able to classify transients down to $r = 22.5$ magnitudes (AB) and, over five years of 4MOST operations, obtain spectra for up to 30 000 live transients to redshift $z \sim 0.5$, measure redshifts for up to 50 000 transient host galaxies to $z \sim 1$ and monitor around 700 active galactic nuclei to $z \sim 2.5$.

## Scientific context

The next decade will see an unprecedented sampling of the extragalactic time-domain universe via massive photometric surveys of the sky. Follow-up spectroscopy of photometric detections is critical to extracting the full astrophysical detail of the objects discovered: their classifications, chemistry, distances (redshifts), luminosities, energetics — and ultimately their physical natures. TiDES addresses this spectroscopic challenge with 250 000 fibre-hours of spectroscopy of transients, their host galaxies, and active galactic nuclei (AGN). These measurements will allow TiDES to tackle three key science goals.

The first goal is the nature of dark energy. This is one of the most puzzling problems in physics, and studying dark energy is the goal of major ground- and space-based facilities over the next decade, for example, Euclid, the Large Synoptic Survey Telescope (LSST) and the Wide Field Infrared Survey Telescope (WFIRST). Type Ia supernovae (SNe Ia) provide a mature and well-exploited probe of the accelerating universe (for example, DES Collaboration et al., 2018), and their use as standardisable candles is an immediate route to measuring the equation of state of dark energy. LSST[1], for example, could assemble around 100 000 SNe Ia[a] to $z = 1$, giving unprecedented insight into the expansion history of the Universe. But a major systematic uncertainty will be the photometric classification and redshift measurement of the supernova detections.

The second goal is to study the extragalactic transient universe. The extragalactic time-domain universe is a far more diverse environment than was imagined only a decade ago. New "superluminous supernovae", "calcium-rich transients", exotic thermonuclear explosions, and even the newly-discovered kilonovae (Smartt et al., 2017) have all demonstrated how little is known about explosive transient populations. LSST will enlarge all these populations by many orders of magnitude and likely uncover entirely new forms of explosions. The key to studying all of these classes of objects is spectroscopy that is rapidly prioritised, which we will implement in TiDES.

The third goal is cosmology and galaxy evolution with AGN. AGN are the most energetic sources in the Universe, showing variability at all wavebands as mass is accreted onto supermassive black holes in the centres of galaxies. The variability of the optical continuum light from the accreting matter and its delayed response mirrored in the optical emission lines of the surrounding material, can be (i) turned into a standard candle similarly to SNe Ia, but out to higher redshifts (Watson et al., 2011) and (ii) used to directly measure the masses of the black holes (for example, Shen et al., 2016). TiDES will establish a Hubble diagram of AGN between $0.1 < z < 2.5$, providing an independent standard candle and delivering the largest catalogue of dynamically measured black hole masses on cosmological scales as a new basis for understanding galaxy evolution.

The majority of TiDES targets will come from LSST, which will produce millions of transient alerts and photometric data on hundreds of thousands of SNe and other variable objects. TiDES will exploit the fact that wherever 4MOST points in the extragalactic sky there will be known time-variable sources, both recently discovered transients, and older, now faded events. Around 30 low-resolution spectrograph (LRS) fibres (2% of the total) in every pointing will be allocated to extragalactic transients, their host galaxies, and AGN. TiDES will therefore "piggyback" on the general 4MOST survey strategy (Figure 1) and will not normally drive the pointing of 4MOST. In addition, 4MOST will regularly (twice per lunation, when schedulable) observe the four announced LSST Deep Drilling Fields (DDFs). These fields are also planned to be observed by the 4MOST WAVES survey (see Driver et al., p. 46).

## Specific scientific goals

### i. Spectroscopic classification of live transients (TiDES-SN)

The aim of TiDES-SN is to observe live transients discovered by LSST and other transient surveys as soon as feasible after discovery. The science goals for TiDES-SN include (i) classification of live SNe, including uncovering rare and unusual events, and (ii) construction of an optimised training sample for the photometric classifiers that will be used to assemble the next generation of SN Ia cosmological samples.

The combination of LSST discoveries and fast turnaround spectroscopic data from TiDES-SN will naturally provide spectroscopic follow-up in the first days after explosion for large samples of transients, including both classical SN types, and more exotic events. Early-time observations (< 3–4 days from transient detection) will allow new insights into the explosion environments and outer layers of the SN ejecta. Rare, fast transients that rise and fall rapidly and that could make important contributions to galactic chemical evolution (for example, calcium-rich fast transients), will also be explored statistically for the first time. We will also target potential lensed SNe that fall within our survey fields (for example, Goldstein et al., 2018) and other transients such as tidal disruption events.

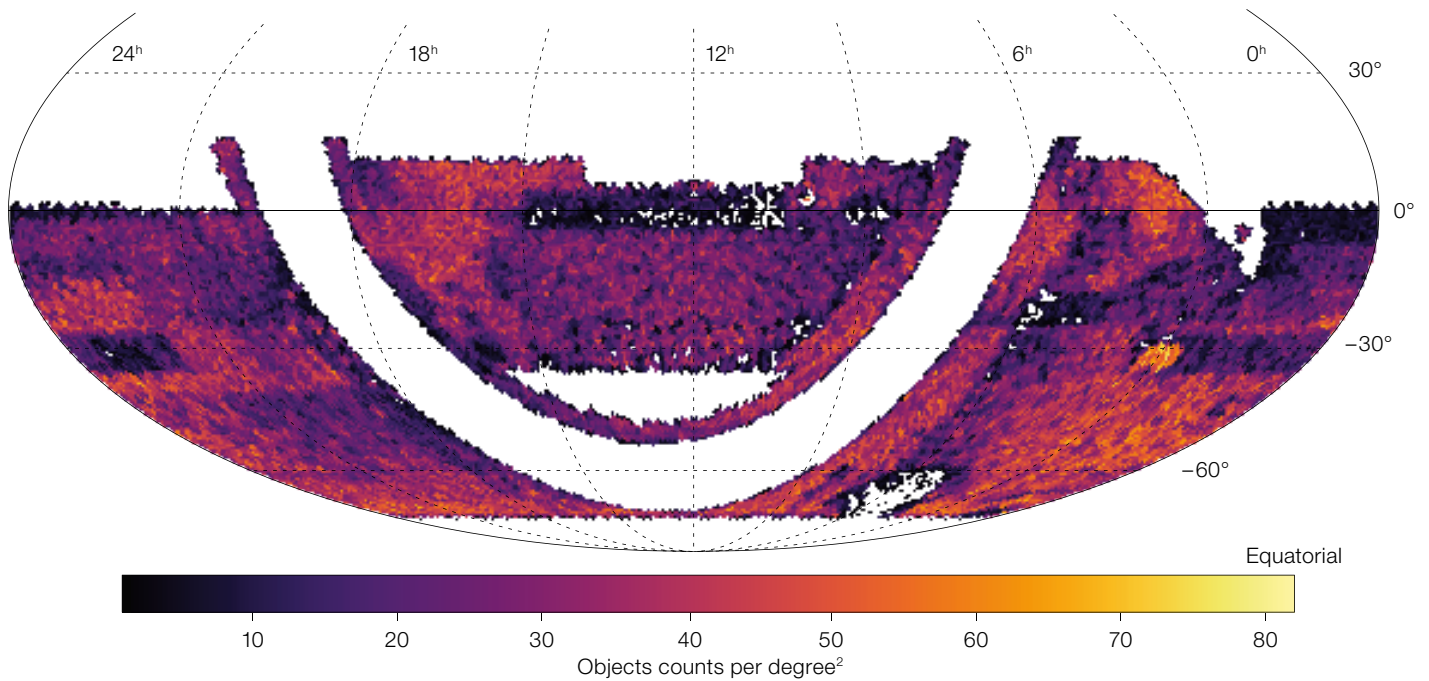

Figure 1. An example distribution of TiDES targets using an example LSST-discovered supernovae distribution as input. The colour scheme shows the typical number[a] of supernovae and their host galaxies per square degree that could be targeted by TiDES. TiDES will optimise the relationship between the two surveys and ensure LSST transients are in the 4MOST queue.

For SN Ia cosmology, even with TiDES, spectroscopic resources are not available to target all candidate SNe Ia, and photometric classification techniques are therefore critical for future SN Ia cosmological analyses. But even the most advanced machine-learning classification techniques are fundamentally dependent on large, homogenous and representative training samples (Lochner et al., 2016). Although TiDES-SN cannot provide a complete sample of SNe, it can provide an unbiased sampling of the whole SN population down to $r = 22.5$ (AB) magnitudes. When combined with the optical (LSST) light curves, this will provide an unsurpassed training sample for future photometric classifiers.

ii. Spectroscopy of supernova host galaxies (TiDES-Hosts)

With TiDES, we will obtain spectroscopic redshifts for host galaxies of SNe that have faded away. This will provide: (i) the SN redshifts required for LSST SN Ia cosmology (see the LSST Dark Energy Science Collaboration Document by Mandelbaum et al., 2018); (ii) improvements in SN photometric classification via the use of spectroscopic redshift priors in the classification algorithms; and (iii) detailed spectral information on the brighter host galaxies, such as metallicity and star formation rates. 4MOST should reach $r = 22.5$ magnitudes in two hours for galaxy redshifts. This limit will be fainter in the deep fields, where our repeat observations will allow the stacking of many hours of spectra. Our model is the Australian Dark Energy Survey (OzDES), a programme at the Anglo-Australian Telescope using the AAOmega spectrograph with the 2dF multi-object fibre positioner, which conducted deep spectroscopic observations of the host galaxies of SNe discovered in 27 square degrees of imaging from the Dark Energy Survey (Childress et al., 2017). 4MOST is expected to obtain host galaxy redshifts for at least 10 times as many SNe Ia, discovered out to $z \sim 1$.

iii. Repeat spectroscopic observations of AGN for reverberation mapping (TiDES-RM)

The primary goal of TiDES-RM is to use AGN broad-line lags to build a Hubble diagram out to $z = 2.5$, and to constrain the cosmological equation of state. AGN as standardisable candles are complementary to SNe Ia, with a redshift distribution that extends to higher redshift. In addition, by using two independent standard candles, TiDES will be insensitive to systematic errors in any individual method, thus increasing the reliability of the results. This will be particularly important when constraining the equation of state of dark energy. Our goal is to extend the redshift range of current surveys, and to exceed the state of the art in early 2020 by at least a factor of two for reverberation-mapped AGN (King et al., 2015). This leads us to target around 700 AGN over the redshift range $0.1 < z < 2.5$.

As a secondary science goal, we will be measuring dynamical masses of supermassive black holes in AGN out to $z \sim 2.5$. Black hole mass is a key parameter in understanding galaxy evolution. Most current black hole mass measurements outside the local universe rely on indirect relations between black hole mass and galaxy properties, for example the M-sigma relation (i.e., the correlation the mass of the supermassive black hole and stellar velocity dispersion). These methods are prone to biases depending on the spatial resolution, which becomes increasingly problematic at higher redshifts. Reverberation-mapped black hole masses have become a standard in the local universe, and we will now push this out to the early universe.

### Science requirements

The science requirements for the three sub-surveys are as follows.

TiDES-SN: Our main requirement to deliver our live transient science is a turn-around time from transient discovery to implementation in the 4MOST schedule of < 3–4 days. Other requirements include a knowledge of the 4MOST pointings well in advance (> 7 days) of a field's being observed, facilitating a smooth link with our transient discovery surveys, and allowing us to focus our target selection on transients in defined areas of the sky. We aim for 30 000 live transient observations, generating datasets large enough to construct a meaningful training sample for photometric classifiers, and statistical samples of rare events.

TiDES-Hosts: Wherever 4MOST points, LSST will have previously discovered SNe in the field. We will put a 4MOST fibre on the host galaxy to measure a redshift. Our target is at least 50 000 successful host galaxy redshifts, which will enable the largest sample of cosmological SNe Ia by at least a factor of 10.

TiDES-RM: The reverberation mapping survey is built on repeat observations in pre-defined and well established LSST extragalactic deep fields that are shared among several Galactic and extragalactic 4MOST surveys. We created an AGN mock catalogue based on the redshift-dependent luminosity function of AGN and simulated C IV and Hβ emission line lags based on established lag-luminosity relations. Based on these lags, a five-year survey duration and the 4MOST signal-to-noise (S/N) estimate for grey time, we determine the required cadence of observations and exposure time per epoch to recover lags for at least the targeted 700 AGN. We find that we require one-hour exposures, corresponding to objects with $r < 21$ magnitudes, and a typical cadence for repeat observations of 14 ± 4 days over an observing semester for each LSST Deep Drilling field.

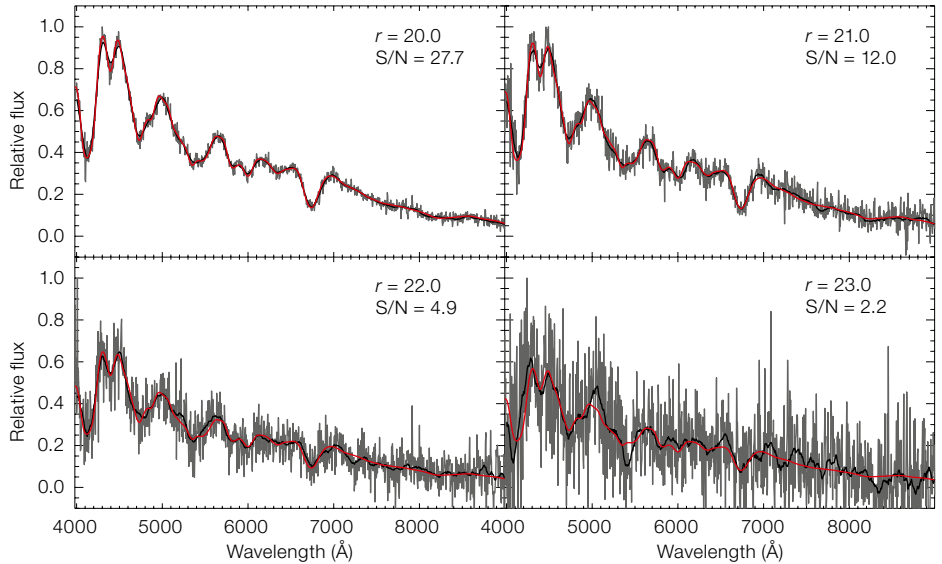

Figure 2. Mock 4MOST spectra based on the 4MOST ETC. Shown here is a maximum-light spectrum of a SN Ia at redshift $z = 0.2$ after "observation" by 4MOST over a series of $r$-band magnitudes, and rebinned to 5 Å (light grey). The mock spectra are calculated assuming an exposure time of 3 × 1200 seconds, average conditions (seeing = 1.1 arcseconds, airmass = 1.3) in dark time, and with 2% sky subtraction residuals. The black lines are the (weighted) Savitzky-Golay filtered mock spectra, and the red spectra show the original template for comparison. The mean S/N per 15 Å bin over 4500–8000 Å is also given.

### Target selection and survey area

TiDES will target the extragalactic fields of major time-domain experiments in the southern hemisphere. We anticipate our principal source of targets to be LSST. This will include both the wide-area LSST deep-wide-fast survey (particularly those areas that have a "rolling cadence"), as well as the narrow-area DDFs. We also require repeat observations of these DDFs. At the time of writing, the LSST observing strategy and survey area have not been finalised, particularly with regard to cadence optimisation, and this will evolve over the coming months and years. As a result, the exact numbers in this section should only be regarded as indicative.

The current TiDES-SN strategy is to target all bright ($r < 22.5$ magnitudes) live transients in a 4MOST field. We expect there to be of order 5–10[a] such transients in any 4MOST extragalactic field on any one night. Over five years with 180 dark/grey nights for extragalactic observations per year, this equates to ~ 30 000 live transients. These targets will be selected in a "blind" fashion, i.e., TiDES-SN will not use colour or other cuts to preferentially select from these objects. However, we will use colour selection or contextual information to prioritise additional interesting, unusual or very early transients that may be fainter than $r = 22.5$ magnitudes.

TiDES-Hosts will preferentially select SNe with a full (LSST) light curve. The number of such hosts will increase with time as LSST builds up the variability history for each field. Simulations using the exposure time calculator (ETC) show that, for a typical SN Ia host galaxy, we expect to be able to obtain a redshift from a spectrum with S/N > 2 per Å. We antici-pate successful redshift measurements for galaxies approximately 2 magnitudes fainter after 10 hours of exposure time before reaching the Poisson limit, a technique adopted by OzDES (Childress et al., 2017).

TiDES-RM will pre-select targets from known AGN in the LSST Deep Drilling fields with a final selection based on their variability history in LSST. The LSST commissioning data preceding 4MOST will be used, with semi-annual updates of monitoring catalogues as LSST builds up a longer time baseline. We will then take one initial spectrum for each candidate AGN to assess whether the emission line profile is suitable for reverberation mapping. If broad lines are detected with S/N > 10 per 15 Å bin in a single 1-hour exposure, the AGN will be monitored until a lag can be determined (depending on luminosity and redshift, this may take between six months and five years), at which point, if feasible, they will be replaced by new sources to maximise the number of measured lags. We require repeat observations of typically one hour, with a cadence from 14 days to 42 days, depending on redshift and luminosity. Based on a mock catalogue selected from the OzDES quasar sample (Tie et al., 2017) of AGN with $r < 21.0$ magnitudes, the expected total execution time will be approximately 25 000 fibre-hours. We note that a sizeable fraction of the AGN monitored by TiDES are also likely targets of the 4MOST AGN survey (S6; see Merloni et al., p. 42).

### Spectral success criteria and figures of merit

Our qualitative spectral success criteria (SSC) are: a successful transient classification (TiDES-SN); a successful redshift measurement (TiDES-Hosts); and a successful AGN spectrum taken (TiDES-RM). The first two criteria are difficult to quantify.

For TiDES-SN, the success will depend on the S/N in the 4MOST spectrum, the transient type (and hence spectral features), and the amount of "contaminating" light from the transient host galaxy. Supernova spectra are dominated by broad features many tens of Angstroms wide, and thus our SSC are defined using S/N in 15 Å bins. Our criteria are based on earlier studies of high-redshift SNe Ia (Balland et al., 2009), where a robust classification can be achieved with a mean S/N = 10 per 15 Å over 4500–8000 Å in the observed frame. This is a conservative success criterion; Balland et al. (2009) demonstrate probable classifications of transients with a mean S/N = 3 per 15 Å. We have also conducted initial simulations using the 4MOST ETC to check our likely classification limits for SNe Ia. We use the ETC to output mock "observed" SN Ia spectra (Figure 2) and attempt a classification with the machine-learning SN classification tool DASH (Muthukrishna et al., in preparation). DASH can classify, with 95% confidence, an $r = 22.2$ magnitudes SN Ia ($z \sim 0.45$) with spectra of S/N = 6 per 1.5 Å bin without using the host redshift as a prior. These simulations were performed with SN spectra free of host galaxy contamination and with "perfect" sky subtraction.

For TiDES-Hosts, we require a S/N ≥ 3 per Å over 4500–8000 Å based on the OzDES survey, which has obtained more than 1700 redshifts with a 95% confidence level of host galaxies with this S/N. Based upon the OzDES project we expect a redshift success rate of approximately 70% for a galaxy with magnitude brighter than $r = 24$ magnitudes in a two-hour exposure. For TiDES-RM, spectral success is defined by achieving a S/N = 10 per 15 Å bin for an AGN spectrum within a predefined cadence.

Our figures of merit (FoMs) encapsulate our broad goals — as many observations of transients and their hosts, and as many time lags measured as possible, for use in astrophysical and cosmological analysis. Our total FoM function is a weighted sum of the three sub-surveys. For TiDES-Hosts and TiDES-SN, the SSC is represented by the error function, with a dependence on the number of objects targeted by 4MOST. For TiDES-RM, the FoM is equal to $x^{1.7}$, where $x$ is the ratio of successfully observed AGN epochs divided by the total number of AGN epochs requested. This functional form captures the fact that no time lag can be determined with fewer than ~ 50% of the requested epochs being successfully observed. While technically this applies for each individual source, the survey average is a good measure of the typical number of epochs observed for a source in the survey.

#### Acknowledgements

TiDES acknowledges funding from Queens University Belfast, Lancaster University, the University of Portsmouth and the University of Southampton.

#### References

Balland, C. et al. 2009, A&A, 507, 85
Childress, M. et al. 2017, MNRAS, 472, 273
DES Collaboration et al. 2018, ApJL, arXiv:1811.02374
Goldstein, D. A. et al. 2018, ApJS, submitted
King, A. L. et al. 2015, MNRAS, 453, 1701
Lochner, M. et al. 2016, ApJS, 225, 31
Mandelbaum, R. et al. 2018, arXiv:1809.01669
Shen, Y. et al. 2016, ApJ, 818, 30
Smartt, S. J. et al. 2017, Nature, 551, 75
Watson, D. et al. 2011, ApJ, 740, 49

#### Links

[1] The Large Synoptic Survey Telescope: https://www.lsst.org/

#### Notes

[a] The exact number will depend on the final LSST observing strategy and implementation.